# Growth and performance of the periodic orbits of a nonlinear driven oscillator

D. R. de Lima[1] and I. L. Caldas[2]

[1] Institute of Mathematics and Statistics, University of São Paulo, São Paulo, Brazil
[2] Institute of Physics, University of São Paulo, São Paulo, Brazil

e-mails addresses: danilolima_bmat@usp.br (D. R. de Lima), ibere@if.usp.br (I. L. Caldas).

**Abstract:** Periodic orbits are fundamental to understand the dynamics of nonlinear systems. In this work, we focus on two aspects of interest regarding periodic orbits, in the context of a dissipative mapping, derived from a prototype model of a non-linear driven oscillator with fast relaxation and a limit cycle. For this map, we show numerically the exponential growth of periodic orbits quantity in certain regions of the parameter space and provide an analytical bound for such growth rate, by making use of the transition matrix associated with a given periodic orbit. Furthermore, we give numerical evidence to support that optimal orbits, those that maximize time averages, are often unstable periodic orbits with low period, by numerically comparing their performance under a family of sinusoidal functions.

## 1. Introduction

Periodic orbits have ever been considered a fundamental knowledge to understand dynamical systems [1]. To investigate properties of these orbits, some maps have been considered, among them the logistic [2], Hénon [3], and the circle map [4].

In fact, the knowledge of the behavior of the periodic orbits of a map can provide a lot of dynamically meaningful information, such as the general structure of the system's attractors [5] and its optimizing invariant measures [6]. Since this realization, many attentions were drawn into this subject. In particular, the simplest question one can ask about a map concerning its periodic orbits is how their quantity grows, as the period gets larger. In this matter, many techniques were developed to count the number of periodic orbits of a given period for certain maps, even obtaining its exact number as the power series expansion of a certain function that depends on the periodic attractors [7].

More recently, questions were posed, also for maps, about the optimality of periodic orbits [8,9,10]. Such topic concerns with the time average of a determined orbit under a measurable function, and what can be said, in general, about the orbits that maximize this average [9]. Those investigations provide many applications. For example, it was shown that one can stabilize a system through small changes near some unstable periodic orbit [11], and that that change gets smaller the less unstable the orbit is. The measure of instability is given by the orbit's Lyapunov exponent [1], which is an example of quantity obtained by the average, over an orbit, of a real valued function.

For the analysis of the periodic orbits, we have chosen a one-dimensional two-parameters circle map. This kind of circle map occurs in a wide class of models, such as the forced Brusselator [12], some electronic oscillators [13,14], and other systems in engineering [15,16] and physics [17,18]. Mathematically, it rises naturally as the discrete representation of a harmonic oscillator perturbed by impulsive kicks at the limit of high dissipation [19]. The considered map has diverse and rich dynamics, as descripted in [20,21] and [22] yet their structure allows us to extract some analytical results.

In this work, we apply the transition matrix technique, along with fundamental results on one dimensional dynamics [23], to estimate the growth number of the unstable periodic orbits of the considered dissipative circle map. Furthermore, trough numerical simulations, we identify some patterns in the evolution of optimal orbits under a family of sinusoidal functions. Thus, we find numerical evidence that optimal orbits are usually unstable periodic orbits with low periods.

In section 2 we introduce the map analyzed in this work. The counting of unstable periodic orbits is in section 3. Identification of optimal period orbits are presented in section 4. The conclusions are in section 5, and some technicalities needed for the main results are left in the Appendices.

## 2. The Dissipative Circle Map

The considered dissipative circle map, comes up as a natural discrete representation of the periodically perturbed system of differential equations

$$\begin{cases} \dfrac{dx}{dt} = -y + sx(1 - x^2 - y^2) + 2a \sum_{n \in \mathbb{Z}} \delta(t - 2\pi b n) \\ \dfrac{dy}{dt} = sy(1 - x^2 - y^2) + x \end{cases} \quad (2.1)$$

in the limit of high dissipation [1, 13]. The perturbation is represented by the Dirac function $\delta$, with intensity $a$ and period $2\pi b$. Without this disturbance, the system is integrable and has very simple dynamics: apart from the unstable singularity in $(0,0)$, every initial condition is attracted to the unitary circumference.

The discrete map generated by (2.1) in polar coordinates is

$$(r_{n+1}, \theta_{n+1}) = \left( \dfrac{1}{1 + (r_n^{-2} - 1)e^{-4\pi sb}} + \dfrac{4a \cos(\theta_n + 2\pi b)}{\sqrt{1 + (r_n^{-2} - 1)e^{-4\pi sb}}} + 4a^2, \arctan\left( \dfrac{\sin(\theta_n + 2\pi b)}{\cos(\theta_n + 2\pi b) + 2a\sqrt{1 + (r_n^{-2} - 1)e^{-4\pi sb}}} \right) \right) \quad (2.2)$$

Letting $s$, the parameter of dissipation, go to infinity, the map converges to the system

$$(r_{n+1}, \theta_{n+1}) = \left( \sqrt{(1 + 4a \cos(\theta_n + 2\pi b) + 4a^2)}, \arctan\left( \dfrac{\sin(\theta_n + 2\pi b)}{\cos(\theta_n + 2\pi b) + 2a} \right) \right) \quad (2.3)$$

which is independent of $r$, allowing the dynamics to be represented by a one-dimensional map in $\theta$:

$$\tan(\theta_{n+1}) = \dfrac{\sin(\theta_n + 2\pi b)}{\cos(\theta_n + 2\pi b) + 2a} \quad (2.4)$$

where $\theta_{n+1}$ is chosen in such a way that $\sin \theta_{n+1} \cdot \sin(\theta_n + 2\pi b) \geq 0$.

When we refer to a difference equation, such as (1.4), as a map, we mean the unique function $f_{a,b}: S^1 \to S^1$ such that $\theta_{n+1} = f_{a,b}(\theta_n)$.

This map, beyond emerging naturally from the perturbed system (2.1) at the limit for $s \to +\infty$ [19], describes oscillations of a coupled oscillator [20]. From a purely mathematical point of view, we regard $f_{a,b}: S^1 \to S^1$ as a continuous map of the circle, for each $a$ and $b$ fixed.

We shall consider the parameters $a$ and $b$ varying in the rectangle $[0,1] \times [0, 0.5]$ as, up to coordinate changes, it represents every possible dynamical behavior observed in the whole plane [20]. Depending of the parameters $a$ and $b$, the map has different solutions such as periodic, quasi-periodic and chaotic attractors. In Fig. 1 are represented the attractors obtained in the parameter space, determined numerically by analyzing the behavior of a typical orbits after neglecting the transient. The colors indicate the attractor observed, as it was reported in [22].

Furthermore, the map has other solutions corresponding to unstable periodic orbits that we analyze in this article, with the goal of determining its asymptotic growth.

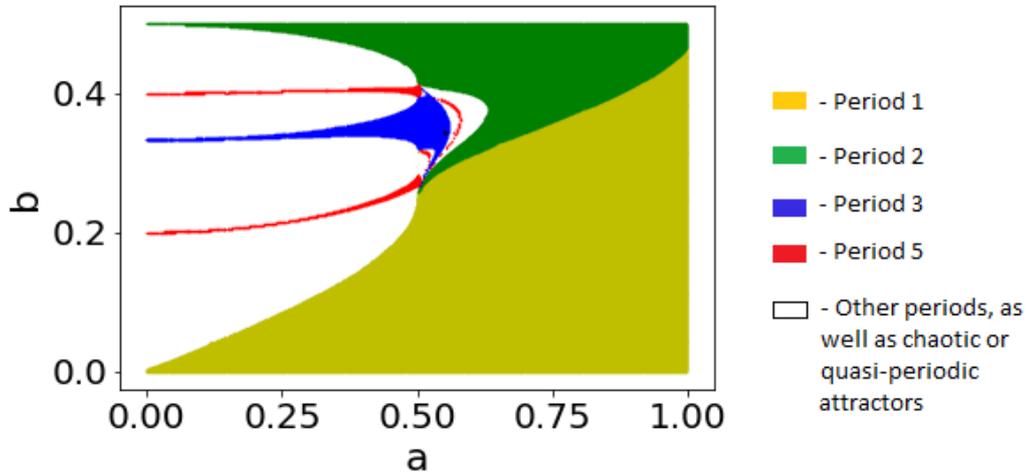

*Figure 1 - Parameter space of the family of dissipative circle maps splitted into different periodic attractors*

For parameters in which $\cos(\theta + 2\pi b) + 2a$ is never zero, $f$ can be extended continuously as a map of the interval $[-\pi, \pi]$, feature that will be crucial to count the unstable periodic orbits. For the remaining parameters, $f$ is a circle homeomorphism, hence, the well-known result that we shall recall in section 3 guaranties that the asymptotic growth of periodic orbits is null [23].

In the next sections, we will continue our analysis of the dissipative circle map under the perspective of counting unstable periodic orbits and determining optimal trajectories [10]. The latter concerns with the time average of a determined orbit under a measurable function, and what can be said, in general, about the orbits that maximize this average [9].

## 3. Counting Unstable Periodic Orbits

A great deal of information can be extracted from a system only by knowledge of the behavior of its stable and unstable periodic orbits [11]. This and the next section will be devoted to their study for certain parameters of the dissipative circle map. We shall start by stablishing some basic nomenclature.

For a map $f$, we will denote by $P(k)$ [resp. $N(k)$] the number of period-$k$ points [resp. orbits] of $f$, and by $F(k)$ the number of fixed points of the iterate $f^k$. We'd also like to distinguish the unstable period-$k$ points, denoting its quantity by $U(k)$.

The following identities are a direct consequence of the above definition:

$$N(k) = \frac{P(k)}{k} \quad (3.1)$$

$$P(k) = F(k) - \sum_{i|k, i \neq k} P(i) \quad (3.2)$$

Our focus is on the asymptotic growth of the previously defined functions, in the following sense:

Let $S(k)$ be a sequence with a finite number of zero terms, we define the *asymptotic growth* of $S(k)$ by:

$$\limsup_{k \to +\infty} \frac{\log(S(k))}{k}$$

if such limit exists. If $S(k) \to 0$ as $k \to +\infty$, we define the asymptotic growth to be zero.

The main method applied in this section concerns estimates of $F(k)$. For that matter, we affirm that for any map $f$, $F(k)$ and $P(k)$ have the same asymptotic growth. The idea of the proof of this fact is to bound the sum in equation (3.2) by the sum over all $i \leq \frac{k}{2}$. The technical details are omitted (see Appendix A).

It's a well-known result that if $f$ is a circle homeomorphism, then the periods of the orbits are bounded [23]. This result classifies half of our maps with respect to periodic orbit counting: Indeed, if $|a| < 0.5$, then since $\cos(\theta + 2\pi b)$ takes all the values in $[-1,1]$, for some $\theta$, $\cos(\theta + 2\pi b) + 2a = 0$.

As we observed, for such parameters, $f_{a,b}$ is a circle homeomorphism and the previous result can be applied, giving that the number periodic orbits of the corresponding map $f_{a,b}$ have zero asymptotic growth for each $b \in [0, 0.5]$.

Another well-known theorem [23] states that for a $C^2$ one-dimensional map with non-flat critical points, there must be only a finite number of attracting periodic orbits. In the notation previously stablished, this means $U(k)$ and $P(k)$ have the same asymptotic growth. Along with the result relating $P$ to $F$, all of the relevant information is encoded by the sequence $F(k)$.

Although we cannot derive a general result valid for any $a > 0.5$, we use the fact that the map is continuous in the interval to use techniques, applicable to particular values of $a$, to obtain upper and lower bounds to the asymptotical growth of $F(k)$.

The upper bound is given by the fact that for those maps, each point have at most two pre-images. Since the critical points of $f^k$ are the pre-images by the some $f^i$, $i \leq k$, of the original critical points, denoting its number by $C(k)$, we obtain inductively

$$C(k) \leq \sum_{i=1}^{k+1} 2^i \leq 2^{k+2} \quad (3.3)$$

We conclude with the observation that between two unstable periodic points of period-$k$, there must be either a stable fixed point or a critical point of $f^k$, and, since the first is finite, we can take $k$ sufficiently large and consider only the later; giving us

$$U(k) \leq C(k) + 1 \leq 2^{k+2} + 1 \quad (3.4)$$

Therefore, the asymptotic growth of $U(k)$ is less or equal than $\log 2$, for any map $f_{a,b}$ with $a > 0.5$.

For the lower bound, we'll start to restrict to particular cases. In order to explain the technique, let's focus on the map with parameters $a = 0.55$ and $b = 0.343$, that we will denote simply by $f$. The main hypothesis that this map provides is the existence of a period-3 orbit. Beyond guarantying the existence of orbits for all periods, via Sharkovskii's Theorem [23], we can infer something about its growth, by a simple step by step construction:

First, define the partition of the periodic orbit to have the periodic points as extremities. Then, construct the associated transition graph, linking an interval $I_i$ to the interval $I_j$ if and only if $f(I_i) \supseteq I_j$. The information of this graph can be encoded in a $2 \times 2$ matrix, defining $a_{ij}$ to be 1, if there's a link from $I_i$ to $I_j$, or 0, otherwise. Denoting the matrix by $M_f$, it has the important property

$$M_{f^k} = (M_f)^k \quad (3.5)$$

where $a_{ij} = k$ means there's $k$ ways of getting from $I_i$ to $I_j$ in $k$ steps.

In particular, if $a_{ii} = k$, then $I_i$ have $k$ fixed points of $f^k$ and, summing over all $i$, that gives the lower bound

$$\log \frac{F(k)}{k} \geq \frac{\log\left(tr(M_f^k)\right)}{k} \quad (3.6)$$

This inequality holds for any continuous interval map and can be applied to any of its periodic orbits. To the special case $a = 0.55$ and $b = 0.343$, the construction is represented in Figure 2, and the right-hand side of (3.6) can be calculated using basic Linear Algebra techniques (see Appendix B), to conclude that the asymptotic growth $h$ is between $\log\left(\frac{1+\sqrt{5}}{2}\right)$ e $\log 2$.

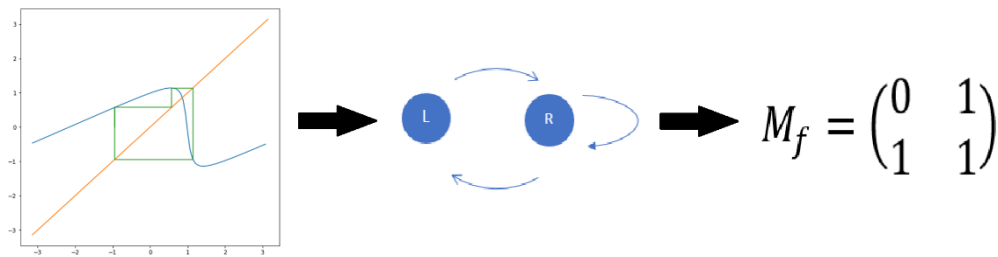

Figure 2 - The transition graph and matrix constructed from the period-3 orbit, shown on the left.

We then conclude that

$$\log\left(\frac{1+\sqrt{5}}{2}\right) \leq h \leq \log 2 \quad (3.7)$$

where $h$ is the asymptotic growth of the map $f$.

Furthermore, a numerical estimative, as illustrated in Fig. 3, where we directly locate and count the periodic points, gives us $h \approx 0.49$, meaning that the asymptotic growth of $P(k)$ is given approximately by

$$P(k) \sim e^{0.49\,k} \quad (3.8)$$

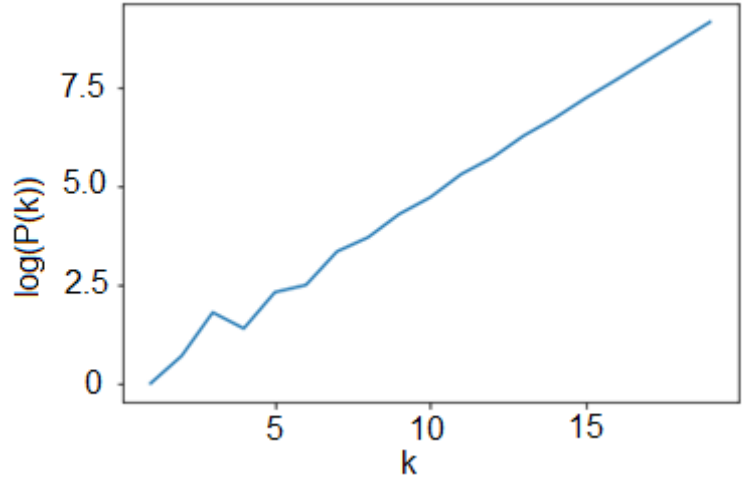

Figure 3 - Numerical calculation of the number of periodic points for $a = 0.55$ and $b = 0.343$, $k \leq 19$ in logarithmic scale.

Thus, we prove that the growth of unstable periodic orbits is exponential and provide an estimative for its rate.

Now we would like to extend this estimative to the largest possible region of the parameter space. We immediately see that this can be done to the part of the period-3 regime, in Figure 1, such that $a > 0.5$ (for those parameters the considered map is continuous and, therefore, the transition matrix method can be applied) consisted of points to which we can trace a path inside the periodic regime area from $(0.55, 0.343)$ (in other words, the path-connected component of the period-3 regime containing $(0.55, 0.343)$).

What if we want a similar estimative for regions corresponding to other periodic regimes? We would then have to deal with transition matrices of order $k - 1$, where $k$ is the period. In a general setting, this would be highly unpractical, since there are many matrices corresponding to periodic orbits, some of them very complicated and others simply not providing any positive lower bound.

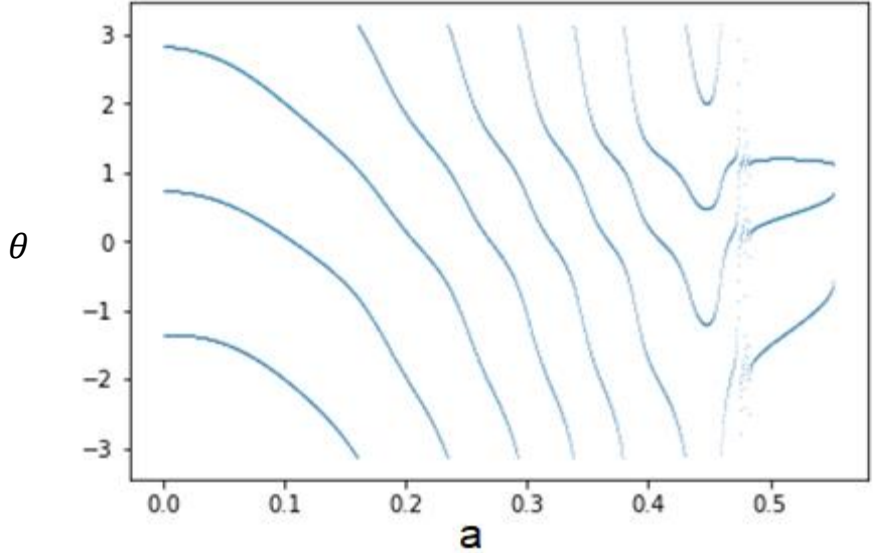

Figure 4 - Bifurcation diagram for $b = \frac{1}{3}$ and $a \in [0, 0.55]$, inside the period-3 stable region, illustrating that the orbit's ordering doesn't change if we identify the upper and lower boundaries of the interval.

However, the key observation we make is that the ordering of the periodic orbit doesn't change up to cyclic permutation inside a periodic regime. This fact can be visualized in Figure 4, where is plotted the bifurcation diagram for $a \in [0, 0.55]$ with $b = \frac{1}{3}$. Notice that, if we glued the upper and lower boundaries of the rectangle together, the period-3 orbit form three continuous paths who never crosses each other. Using this fact, we can make some restrictions. In fact, if the parameter is close enough to the vertical line $a = 0.5$, we can trace it back to a rational rotation (that is, a parameter for which $a = 0$ and $b$ is rational and produces period-$k$ orbits). Therefore, the only matrices we have to analyze are the ones corresponding to a periodic orbit of a rotation. Simplifying even further, we can iterate the rotation $p$ times, for certain $p$, such that, if $\{x_1 ..., x_k\}$ is the periodic orbit in increasing real order, then $f^p(x_1) = x_2, f^p(x_2) = x_3, ..., f^p(x_{k-1}) = x_k$ and $f^p(x_k) = x_1$. The transition matrix associated is the $(k-1)$-th order matrix:

$$M_k = \begin{pmatrix} 0 & 1 & ... & 0 \\ \vdots & 0 & \ddots & \vdots \\ 0 & ... & 0 & 1 \\ 1 & ... & ... & 1 \end{pmatrix} \quad (2.9)$$

That is, the matrix element $a_{ij} = 1$, if $j = i + 1$ or $i = k - 1$, and $a_{ij} = 0$ otherwise. Let $r_k$ be the largest eigenvalue of $M_k$. As computed before, for $k = 3$, it can be shown that $1 < r_k < 2$ and, therefore, $\log r_k$ gives us a positive lower bound for the periodic orbits' growth number (in fact, our previous estimative for $(0.55, 0.343)$ relies on the fact that $r_2 = \frac{1+\sqrt{5}}{2}$).

More than that, it can be proved that those values get closer to $\log 2$, which is the universal upper bound. Therefore, the largest the period of the region, the more precise is the growth number estimative, using this method. More precisely, if $h_k$ denotes the growth number for a parameter $(a, b)$ inside a period-$k$ regime, with $a > 0.5$, and in the same path-connected component of a rotation, then

$$\log r_k \leq h \leq \log 2 \quad (3.10)$$

with $\log r_k \to \log 2$, as $k \to \infty$.

Thus, we extend the initial counting predictions from a particular choice of parameters $a, b$ to parameters in a larger area of the parameter space.

## 4. Optimal Trajectories

Many applications of dynamical systems involve maximizing the average of a real valued function over the orbits of a map. We will call these maximizing orbits "optimal".

In this section, we will make clear the hypothesis over the system and the performance measuring function, briefly recall well-known results of Ergodic Theory and some conjectures about optimal orbits, and finally bring those questions to the specific case of the dissipative circle maps family, identifying patterns through numerical simulations.

The objects we will be dealing with are pairs $(f, \varphi)$, where $f: X \to X$ is a map on a measure space $X$ (in our case, $X$ is the unitary circle $S^1$ or the interval $[-\pi, \pi]$), and $\varphi: X \to \mathbb{R}$ is a measurable function.

Furthermore, our interest is restricted to atypical orbits. By that we mean orbits that generate time averages different from almost every orbit, which generates a default performance.

Not every system makes this restriction possible. Take, for example, a system with two periodic attractors, each with a positive measure basin, providing two different performances. For such a system, we wouldn't have a default performance, since there are two different average values, each one achieved by a positive measure set of initial conditions. Hence the importance of the Birkhoff Ergodic Theorem [10], that states the following:

If $f: X \to X$ have an invariant measure $\mu$ (we may assume $\mu(X) = 1$), then for any $\varphi: X \to \mathbb{R}$ measurable,

$$\lim_{n \to \infty} \frac{1}{n} \sum_{k=1}^{n} \varphi\left(f^k(x)\right)$$

exists and is the same for $\mu$-almost every $x \in X$. By that we mean that if $A$ is the set of points that doesn't satisfy the property, then $\mu(A) = 0$. We call such limit the *time-average* and denote it by $\langle \varphi \rangle(x)$. Furthermore, if $f$ is ergodic with respect to $\mu$, meaning that the only subsets $A \subset X$ satisfying $f^{-1}(A) = A$ must have $\mu(A) = 0$ or $\mu(A) = 1$, this average equals the spatial average, that is

$$\langle \varphi \rangle(x) = \int_X \varphi(x) \, d\mu(x) \quad (4.1)$$

This reduces the task of checking the limit for every initial condition to prove ergodicity, for which there are several techniques.

In the following, we run some numerical simulations to see how the family of circle maps behaves with respect to optimal trajectories. The family of real valued functions is chosen to be

$$f_\gamma(x) = \cos(x - 2\pi\gamma)$$

as is usual in the literature. The parameters chosen are $a = 0.57$ and $b = 0.343$.

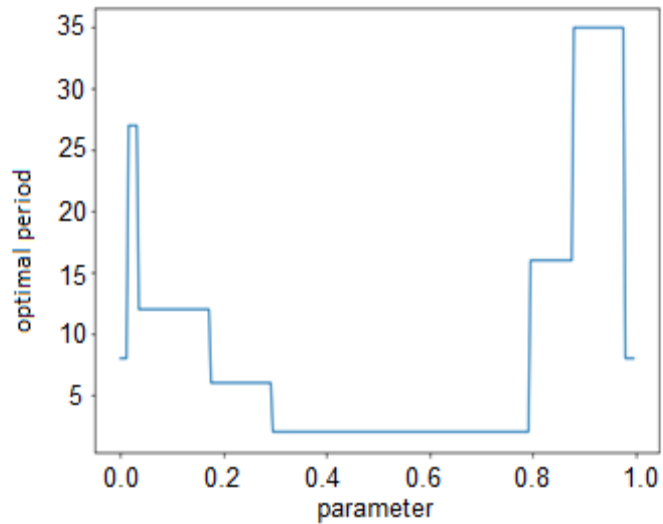

Figure 5 – Optimal periods for the family of functions $f_\gamma(x) = cos(x - 2\pi\gamma)$, for each $\gamma \in [0,1]$.

A more insightful way to look at the information of Figure 5, at least towards the conjectured in [9], is displayed in Figure 6, where we consider how the probability of a period-$p$ orbit being optimal decays with increasing $p$.

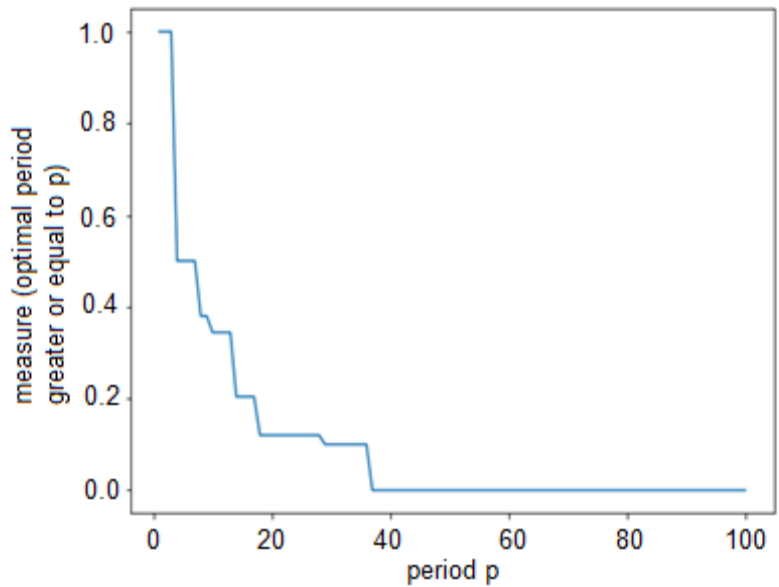

Figure 6 – For each $p \leq 100$, we plot the measure of the set of parameters $\gamma$ that produce an optimal period greater or equal to $p$.

The literature on ergodic optimization usually talks about maximizing measures, rather than orbits. So, in order to discuss the currently known results, it is important to explain the link between the concepts.

To every initial condition $x_0$ of a map $f$ can be assigned a natural measure, given by

$$\mu_{x_0} = \lim_{n \to \infty} \frac{1}{n} \sum_{i=0}^{n-1} \delta_{f^i(x_0)}$$

where, for a point $y$ and a set $A$, $\delta_y(A) = 1$ if $y \in A$ and $0$ if $y \notin A$. This measure represents the proportion of the orbit inside the set $A$ and it is called the *measure generated by the orbit of $x_0$*. The time average of an orbit under a performance function can also be given in terms of the generated measure, through the following formula:

$$\langle \varphi \rangle(x_0) = \int_X \varphi(x) \, d\mu_{x_0}(x) \quad (4.2)$$

Then, is natural to switch the focus to the question of what invariant measures maximizes the time average, since those encapsulate the ones generated by orbits.

We call the support of a probability measure $\mu$ the smallest closed set $F$ such that $\mu(F) = 1$. In this language, the question proposed in [9] becomes: when is an optimizing measure supported on an orbit, and how often is this orbit periodic? In this direction, it was suggested that natural measures can be obtained in chaotic systems from the analysis over the unstable periodic orbits [24].

Although it is possible to provide pathological examples of one-parameter families for which all optimal orbits are non-periodic, we believe, based on the numerical simulation in simpler cases, that simple additional hypotheses concerning how the curve is distributed in function space should be helpful in the path to answering that question (see Appendix C).

In this section we gave numerical evidence for a circle mapping that the optimal orbit, for generic smooth functions, is typically a periodic orbit of low period, as it was conjectured in [9]. However, it was pointed out in [25] that this may not be valid generally for continuous system, once optimal time averages may be achieved by long-period unstable orbits. In fact, in [25] it was presented an example of optimal long-period orbits that spend substantial amounts of time in a region of phase space that is close to large values of the performance function. However, an essential difference for flows is that chaotic attractors can have embedded within them, not only unstable periodic orbits, but also unstable steady states, and optimality can often occur on steady states [26]. Nonetheless optimality is typically achieved at low period [26]. Furthermore, optimal periodic orbits are insensitive to small perturbations of a smooth performance function of the system state, while the optimality of a non-periodic orbit can be destroyed by arbitrarily small perturbations [27]. Complementary, only a few unstable orbits with low periods give good mean statistical properties in dynamical systems in fluid dynamics [28].

## 5. Conclusion

In this article we analyzed properties of periodic orbits of a dissipative circle map, obtained from a system of differential equations with a stable limit cycle and an unstable fixed point in the center. The analyzed map was useful to apply our methods of counting the number of periodic orbits and determining the main characteristics of the optimal trajectories.

We illustrated how the asymptotic growth of periodic orbits of a system can be extracted from very little information. In this case, just the existence of a low period orbit and the way it is ordered allowed us to construct the transition graph and matrix, through which we estimated the lower bound for the growth number. We also observed that this information is carried through a continuous path inside a periodic regime, allowing the estimative to be extended from one point to a large region.

Furthermore, we've searched for optimal trajectories under a sinusoidal performance function and, through numerical simulations, provided evidence to conclude that optimal trajectories are often periodic ones with low periods. More precisely, the probability of a given unstable periodic orbit being optimal seems to decay exponentially with the period. However, for continuous systems, as mentioned in section 4, even so the optimality is typically achieved at low periods, examples have been found of optimal time averages achieved by long-period unstable orbits.

## Acknowledgments

The authors acknowledge the financial support from the Brazilian Federal Agency CNPq, under Grant Nos. 137141/2020-3 and 302665/ 2017-0, and the São Paulo Research Foundation (FAPESP, Brazil), under Grant No. 2018/03211–6.

## References

[1] K. T. Alligood, T. D. Sauer, J. A. Yorke, "Chaos, an introduction to dynamical Systems" Springer, 1997.

[2] May, Robert M. "Simple mathematical models with very complicated dynamics." Nature 261.5560 (1976): 459-467.

[3] M. Hénon, "A two-dimensional mapping with a strange attractor." The Theory of Chaotic Attractors. Springer, New York, NY, 1976. 94-102.

[4] Jensen, M. Høgh, Per Bak, and Tomas Bohr. "Complete devil's staircase, fractal dimension, and universality of mode-locking structure in the circle map." Physical review letters 50.21 (1983): 1637.

[5] Auerbach, D., Cvitanović, P., Eckmann, J. P., Gunaratne, G., & Procaccia, I. (1987). "Exploring chaotic motion through periodic orbits." Physical Review Letters, 58(23), 2387.

[6] P. Cvitanović, "Invariant measurement of strange sets in terms of cycles." Physical Review Letters 61.24 (1988): 2729.

[7] J. Milnor , and W. Thurston. "On iterated maps of the interval." Dynamical systems. Springer, Berlin, Heidelberg, 1988. 465-563.

[8] O. Jenkinson, "Maximum hitting frequency and fastest mean return time", Nonlinearity, 18 (2005), 2305–2321.

[9] B. R. Hunt, E. Ott, "Optimal periodic orbits of chaotic systems", Phys. Rev. Lett. 76, 2254 (1996)

[10] O. Jenkinson, "Ergodic optimization", Discrete Contin. Dynam. Systems 15 (2006), no. 1, p. 197–224.

[11] T. Shinbrot, C. Grebogi, E. Ott, J. A. Yorke, "Using small perturbations to control chaos", Nature 363, 411 (1993).

[12] B. Hao, S. Zhang, "Hierarchy of chaotic bands." J Stat Phys 28, 769–792 (1982); B. L. Hao, G. R. Wang, and S. Y. Zhang, Commun. Theor. Phys. 2, 1075 (1983).

[13] D. L. Gonzalez, and O. Piro. "Chaos in a nonlinear driven oscillator with exact solution." Physical Review Letters 50.12 (1983): 870.

[14] D. L. Gonzalez, and O. Piro. "One-dimensional Poincaré map for a non-linear driven oscillator: Analytical derivation and geometrical properties." Physics Letters A 101.9 (1984): 455-458.

[15] R. E. Mickens. Oscillations in planar dynamics systems. Singapore: World Scientific (1996).

[16] R. E. Mickens, A. B. Gumel, J Sound Vibr. 250, 955-956 (2002).

[17] J. Hale, Ordinary differential equations, New York: Wiley (1969).

[18] J. Guckenheimer, P. Holmes, Nonlinear oscillations, dynamical systems, and bifurcations of vector fields. New York: Springer-Verlag (1983).

[19] E.J. Ding. "Analytic treatment of periodic orbit systematics for a nonlinear driven oscillator" Phys. Rev., A34:3547-3550, 1986.

[20] K. Ullmann, and I. L. Caldas. "Transitions in the parameter space of a periodically forced dissipative system." Chaos, Solitons & Fractals 7.11 (1996): 1913-1921.

[21] K. Ullmann. Métodos de Análise de Mapeamentos Não-Lineares com Aplicação à Física de Plasmas [thesis]. São Paulo: Instituto de Física; 1998. doi:10.11606/T.43.1998.tde-28022014-115623.

[22] W. Façanha, B. Oldeman, and L. Glass. "Bifurcation structures in two-dimensional maps: The endoskeletons of shrimps." Physics Letters A 377.18 (2013): 1264-1268.

[23] W. de Melo, S. van Strien, One-Dimensional Dynamics. Springer - Verlag, 1993

[24] Lai, Ying-Cheng, Yoshihiko Nagai, and Celso Grebogi. "Characterization of the natural measure by unstable periodic orbits in chaotic attractors." Physical Review Letters 79.4 (1997): 649.

[25] Zoldi, Scott M., and Henry S. Greenside. "Comment on "Optimal Periodic Orbits of Chaotic Systems"." Physical review letters 80.8 (1998): 1790.

[26] Yang, Tsung-Hsun, B. R. Hunt, and E. Ott. "Optimal periodic orbits of continuous time chaotic systems." Physical Review E 62.2 (2000): 1950.

[27] G. Yuan, and B. R. Hunt. "Optimal orbits of hyperbolic systems." Nonlinearity 12.4 (1999): 1207.

[28] Y. Saiki, and M. Yamada. "Time-averaged properties of unstable periodic orbits and chaotic orbits in ordinary differential equation systems." Physical Review E 79.1 (2009): 015201.

## Appendix A – Fixed and Periodic Points Growth

Let $h = \lim\limits_{k \to \infty} \frac{\log(F(k))}{k}$.

We clearly have $P(k) \leq F(k)$. Hence, $\lim\limits_{k \to \infty} \frac{\log(P(k))}{k} \leq h$.

Define $R(k) = F(k) - P(k)$. Equation (2.2) gives us

$$R(k) = \sum_{i | k / i \neq k} P(i)$$

It will suffice to show that $\lim\limits_{k \to \infty} \frac{R(k)}{N(k)} = 0$.

For that we consider the following upper bound for $R$:

$$R(k) \leq \sum_{i=1}^{\lfloor\frac{k}{2}\rfloor} P(i) \leq \frac{k}{2}\max\left\{P(i): i \leq \frac{k}{2}\right\} \leq \frac{k}{2}\max\left\{F(i): i \leq \frac{k}{2}\right\}$$

By hypothesis, $\lim_{k\to\infty} \frac{\log(F(k))}{k} = h$, so $\forall \varepsilon > 0$, there exists $N \in \mathbb{N}$ such that for every $k \geq N$, $\frac{\log F(k)}{k} < h + \varepsilon \Rightarrow F(k) < e^{(h+\varepsilon)k}$.

To conclude the construction of the upper bound, note that for each $\varepsilon > 0$ there is only finite $k$ such that $F(k) \geq e^{(h+\varepsilon)k}$ and, therefore, we can take the maximum, say $m_\varepsilon$. Hence, there is $k_0 \in \mathbb{N}$, (we can take it greater than $N$) such that for larger values of $k$, the function $e^{(h+\varepsilon)k} > m_\varepsilon$. For those values the following holds:

$$R(k) \leq \frac{k}{2} e^{(h+\varepsilon)\frac{k}{2}}$$

That gives us automatically a lower bound for $F(k)$:

$$N(k) = F(k) - R(k) \geq F(k) - \frac{k}{2} e^{(h+\varepsilon)k}$$

Taking $k$ sufficiently large such that $\frac{\log F(k)}{k} > h - \varepsilon$, we obtain:

$$N(k) \geq e^{(h-\varepsilon)k} - \frac{k}{2} e^{(h+\varepsilon)k}$$

Hence:

$$\frac{N(k)}{R(k)} \geq \frac{e^{(h-\varepsilon)k}}{\frac{k}{2} e^{(h+\varepsilon)\frac{k}{2}}} - 1 \geq \frac{2e^{(h-\varepsilon)k - (h+\varepsilon)\frac{k}{2}}}{k} = \frac{2e^{\frac{k}{2}(h-3\varepsilon)}}{k}$$

Since $h > 0$ and the inequalities holds for every $\varepsilon > 0$, we choose $\varepsilon < \frac{h}{3}$, implying $h - 3\varepsilon > 0$. A simple application of the L'Hospital rule gives $\frac{2e^{\frac{k}{2}(h-3\varepsilon)}}{k} \to +\infty$, if $k \to \infty$ and, therefore, $\frac{N(k)}{R(k)}$ goes to infinity as well

To conclude the proof:

$$\lim_{k\to\infty} \frac{\log(F(k))}{k} - \frac{\log(P(k))}{k} = \lim_{k\to\infty} \frac{\log\left(\frac{F(k)}{P(k)}\right)}{k} = \lim_{k\to\infty} \frac{\log\left(\frac{P(k)+R(k)}{P(k)}\right)}{k} =$$

$$\lim_{k\to\infty} \frac{\log\left(1 + \frac{R(k)}{P(k)}\right)}{k}$$

Since $\frac{P(k)}{R(k)} \to +\infty$, we have $\frac{R(k)}{P(k)} \to 0$ and, therefore, the above limit is zero.

Going back to the beginning, we have $\lim_{k\to\infty} \frac{\log(P(k))}{k} = \lim_{k\to\infty} \frac{\log(F(k))}{k} = h$, finishing the proof.

## Appendix B – Asymptotic Growth for $a = 0.55$ and $b = 0.343$

Denoted by $h$ the desired asymptotic growth. We already know that $h \leq \log 2$, since it is a universal upper bound for the family.

Now, we obtain the lower bound using inequality (2.6), where

$$M_f = \begin{pmatrix} 0 & 1 \\ 1 & 1 \end{pmatrix}$$

The eigenvalues of this matrix are the roots of the characteristic polynomial $P(t) = t^2 - t - 1$, $\lambda_1 = \frac{1+\sqrt{5}}{2}$ and $\lambda_2 = \frac{1-\sqrt{5}}{2}$. Thus, we have

$$\text{tr}(M_f^k) = \left(\frac{1+\sqrt{5}}{2}\right)^k + \left(\frac{1-\sqrt{5}}{2}\right)^k$$

and since $\left(\frac{1-\sqrt{5}}{2}\right)^k \to 0$, in the limit we have

$$\lim_{k \to \infty} \frac{\log\left[\left(\frac{1+\sqrt{5}}{2}\right)^k + \left(\frac{1-\sqrt{5}}{2}\right)^k\right]}{k} =$$

$$\lim_{k \to \infty} \frac{\log\left(\frac{1+\sqrt{5}}{2}\right)^k}{k} = \log\left(\frac{1+\sqrt{5}}{2}\right)$$

Finally, taking the limit on both sides of (2.6) gives us $h \geq \log\left(\frac{1+\sqrt{5}}{2}\right)$.

## Appendix C – Optimality Regions

Let $T: X \to X$ be a map, where $X = [-\pi, \pi]$ or $S^1$, and $V$ a vector space of functions $f: X \to \mathbb{R}$ (for example, it could be the space of $k$–Lipchsitz functions, $C^r$ functions, measurable functions or intersection of these spaces, as long as the property that defines it is preserved by linear combination). For abbreviation, let's call a function for which optimality is achieved in a non-periodic orbit atypical.

We can write the space $V$ as the following union over $x \in X$:

$$V = \bigcup_{x \in X} R_x$$

where $R_x = \{f \in V : \langle f \rangle(x) \geq \langle f \rangle(y), \forall y \in X\}$ is the set of functions that make the orbit of $x$ optimal. Notice that those sets are not disjoint, for example any constant function is contained in $R_x$, for every $x$. If we want the disjointness property for points in different orbits, as well as identify strict optimality, we can simply define $R_x^+ = \{f \in V : \langle f \rangle(x) > \langle f \rangle(y), \forall y \in X\}$.

The key observation is that those optimality and strict optimality regions behave simply with respect to linear combinations. The next claim makes that precise: Let $f$ and $g$ be performance functions such that

$$\langle f \rangle(x) > \langle f \rangle(y)$$
$$\langle g \rangle(x) \geq \langle g \rangle(y)$$

Then, for each $\alpha > 0, \beta \geq 0$, defining $h = \alpha f + \beta g$, we have

$$\langle h \rangle(x) > \langle h \rangle(y)$$

This follows from the linearity of the time average and the simple calculation:

$$\langle h \rangle(x) = \langle \alpha b + \beta g \rangle(x) = \alpha \langle f \rangle(x) + \beta \langle g \rangle(x) > \alpha \langle f \rangle(y) + \beta \langle g \rangle(y) = \langle h \rangle(y)$$

As a consequence, if $f$ is atypical, then for $\alpha > 0$ and any constant $c \in \mathbb{R}$, the function $h = \alpha f + c$ is also atypical. To see that, take $x$ as a point of the optimal orbit, $y$ a generic point and $g$ a constant function.

From that, we conclude that for each example of functions yielding optimality at a non-periodic orbit, there is at least a hyperplane of functions with that same property and, therefore, one-parameter families of functions chosen inside these hyperplanes will never make a periodic orbit optimal.

That being said, there are two important points as why the conjectures in [9] can still hold under a mild restriction. First, this is not a constructive result, it depends on first locating examples of atypical functions. Second, since we're talking about spaces of functions, the number of dimensions involved makes it almost impossible, in normal examples, for a one-parameter family to be contained in a two-dimensional space, unless one chose it to be that way (for example, any two parameters in the sinusoidal family considered in section 4 gives two linearly independent functions, therefore the family is not contained in any finite dimensional vector space). In that matter, we believe a "transversality" hypothesis, guarantying that the family doesn't stay inside those atypical spaces too long, should not be very restrictive and yet provide new tools in the direction of giving a positive answer to the conjecture.